\documentclass[a4paper,twocolumn,english,aps,prb,showpacs]{revtex4}
\usepackage{pslatex}
\usepackage[T1]{fontenc}
\usepackage[latin1]{inputenc}
\usepackage{array}
\usepackage{amsmath}
\usepackage{color}
\usepackage{graphicx}
\usepackage{amssymb}

\makeatletter

\providecommand{\tabularnewline}{\\}

\usepackage{babel}
\makeatother
\begin{document}

\title{Local tetragonal distortion in $La_{0.7}Sr_{0.3}MnO_{3}$ strained
thin films probed by x-ray absorption spectroscopy}

\author{Narcizo M. Souza-Neto}

\altaffiliation[Also at: ]{Instituto de Física Gleb Wataghin, IFGW - UNICAMP, Campinas, SP,
Brazil}

\email{narcizo@lnls.br}

\author{Aline Y. Ramos}

\altaffiliation[Also at: ]{Laboratoire de Minéralogie-Cristallographie de Paris, LMCP -UMR 7590
-CNRS, Paris, France}

\author{Hélio C. N. Tolentino}

\affiliation{Laboratório Nacional de Luz Síncrotron - LNLS, P.O. Box 6192, 13084-971,
Campinas, São Paulo, Brazil}

\author{Emmanuel Favre-Nicolin}

\author{Laurent Ranno}

\affiliation{Laboratoire Louis Néel, UPR 5051 CNRS-UJF, Grenoble, France}

\date{\today{}}

\begin{abstract}
We report on an angular resolved X-ray Absorption Spectroscopy study
of $La_{0.7}Sr_{0.3}MnO_{3}$ thin films epitaxially grown by pulsed
laser deposition on slightly mismatched substrates which induce tensile
or compressive strains. XANES spectra give evidence of tetragonal
distortion within the $MnO_{6}$ octahedra, with opposite directions
for tensile and compressive strains. Quantitative analysis has been
done and a model of tetragonal distortion reflecting the strain has
been established. EXAFS data collected in plane for tensile substrate
confirm the change in the $Mn-O$ average bond distance and the increase
of $Mn-Mn$ length matching with the enlargement of the cell parameter.
From these results we conclude that there is no significant change
in the $Mn-O-Mn$ angle. Our observations conflict with the scenarios
which this angle is the main driving parameter in the sensitivity
of manganite films properties to external strains and suggest that
the distortion within the octahedra plays a key role in the modification
of the transport and magnetic properties.
\end{abstract}


\maketitle

\section{Introduction}

The delicate balance between orbital and spin interactions in doped
manganites ($RE_{1-x}A_{x}MnO_{3}$, $RE$ = rare earth, and $A$
= alkaline metal) leads to many interesting exotic properties. Among
them the colossal magneto-resistance brought about the stir of interest
for this class of material \cite{Jin-Science94,Millis-N98,Murakami-N03}.
It has been largely reported that the remarkable properties of doped
manganites show drastic sensitivity to small changes in their structural
parameters and the form of the samples\cite{Mahesh-JSSC95,Mahendiran-PRB96}.
At the atomic level this versatility is correlated to distortion within
the manganese coordination octahedra $MnO_{6}$ and modifications
in the angle $\varphi=Mn-O-Mn$\cite{Fontcuberta-PRL96,Moritomo-PRB95,Jin-Science94}.
The $Mn^{3+}$ ion has an high-spin $t_{2g}^{3}e_{g}^{1}$ configuration
and the deformation of the metal-ligand octahedra stems from the Jahn-Teller
interaction, that tends to lift the degeneracy of the $d$ orbitals
and stabilize the energy levels of the occupied $d$ electron. The
angle $\varphi$ is $\pi$ in a cubic structure, but is bent and deviated
from $\pi$ in non-cubic compounds. Thin films of manganites, grown
using deposition techniques similar to the ones developed for high-temperature
superconductors (pulsed laser deposition, sputtering, MOCVD,$\ldots$
), drawn large possibilities in the design of tunable magnetic devices\cite{Prellier-JPCM01}.
Thus it has been early observed that manganites films display properties
significantly different from those of the bulk material and these
properties are dependent on the film thickness\cite{Millis-JAP98,Prellier-JPCM01,Mahesh-JSSC95,Wang-APL99}.
One prominent example is the decrease of the Curie temperature $T_{C}$
as the strain increases \cite{Ranno-ASS02,Bibes-PRB02}. This characteristic
has been associated to the strain induced by the lattice mismatch.
It should be noted that in bulk manganites the sensitivity of $T_{C}$
for applied hydrostatic pressure and ''chemical'' pressure, i.e.
structural distortion changing the average radius of the atom in the
$A$ site, is well known and associated to modification in angle $\varphi$.
This correlation is simply derived when one remembers that description
of the transport and magnetic properties are determined by the effective
intersite hopping $t_{ij}$ of the $e_{g}$ electrons via O $2p$
states that controls the double exchange. In the strong ligand field
approximation the $p-d$ transfer interaction is scaled as $t_{pd}=t_{pd}^{\circ}cos(\varphi)$,
($t_{pd}^{\circ}$ transfer interaction for $\varphi=\pi$) then $T_{C}$
is almost proportional to $cos^{2}\varphi$ \cite{Radaelli-PRB97,Fontcuberta-PRL96,Moritomo-PRB95}.
As $t_{pd}^{\circ}$ depends on the $Mn-O$ distance, the transport
properties are also expected to vary with this distance. To understand
the correlation between the substrate-induced strain and the modifications
of the transport and magnetic properties in thin films, it is then
important to explicit the connection between the crystallographic
cell strains and the modifications in the local parameters\cite{Millis-JAP98,Fontcuberta-PRL96,Prellier-APL00}.
Several experimental studies by X-ray Absorption spectroscopy have
addressed the local organization in manganites films, and variations
of the local environment in thin films versus bulk samples, pointing
out the importance of the local distortion around $Mn$\cite{Cao-PRB00,Miniotas-JAP01,Qian-PRB01_2}.
At the moment it does not seem that a full consensus is reached about
what the main distortion at the local scale is. In $La_{x}Ca_{1-x}MnO_{3}$
films, with x close to 0.7, Miniotas and coworkers\cite{Miniotas-JAP01}
do not observe any modification within the coordination shell, but
reported on variations of the Mn-Mn distance, according to the substrate
misfit strain. They conclude that the biaxial strain is accommodated
by appropriate bending of the $Mn-O-Mn$ angle. On the other hand
thickness-dependent $Mn$ coordination asymmetry is reported in thin
film of the system $Nd_{0.5}Sr_{0.5}MnO_{3}$ close to the charge-ordered
state\cite{Qian-PRB01_2}. These results suggest the importance of
the biaxial strain on the electron localization, quite different from
the effect of an hydrostatic pressure. 

$La_{0.7}Sr_{0.3}MnO_{3}$ manganite exhibits ferromagnetic transition
around room temperature ($T_{c}\approx360K$ \cite{Urushibara-PRB95,Mahendiran-PRB96,Moritomo-PRB95,Mahesh-JSSC95})
and fully spin-polarized conduction band. These characteristics made
this compound attractive as low cost magnetic sensor and a perfect
model material for correlated electrons properties. $La_{0.7}Sr_{0.3}MnO_{3}$
crystallize in a the rhomboedral $R\bar{3}c$ ($D_{3d}^{6})$ variant
of the cubic perovskite, with $La$ and $Sr$ randomly distributed
on the $A$ site (6a positions) \cite{Radaelli-PRB97,Mahendiran-PRB96}.
The ratio between the cell parameter a and c is close to $1/\sqrt{2}\sqrt{3}$,
so that this structure can be considered as {}``pseudo-cubic'' with
a=3.87$\textrm{Å}$. From the crystallographic data the angles $\varphi$
are close to $165^{\circ}$ and there is no distortion of the $MnO_{6}$
octahedra. A measure of this distortion is given as the root square
deviation of the $Mn-O$ distances from their average value: $\sigma_{JT}=\sqrt{\frac{1}{3}{\displaystyle {\textstyle {\scriptstyle {\textstyle \sum_{i}\left[(Mn-O)_{i}-\left\langle Mn-O\right\rangle \right]^{2}}}}}}$
, which is of about $1.2\times10^{-1}\textrm{Å}$ in the prototype
Jahn-Teller compound $LaMnO_{3}$ at room temperature associated to
the strong localization of the $e_{g}$ electron. In the metallic
$La_{0.7}Sr_{0.3}MnO_{3}$ compound, the $e_{g}$ electron is delocalized
and the contribution of the spontaneous energy-lowering Jahn-Teller
distortion of the $Mn^{3+}O_{6}$\cite{Wu-PRB00,Urushibara-PRB95}
does not give rise to a measurable average distortion of the octahedra.
In X-ray absorption spectroscopy (XAS) the coordination shell around
Mn atoms can be modelled with one unique $Mn-O$ distance at $\approx$$1.94\textrm{Å}$\cite{Mastelaro-XRS02,Shibata-PRB03}.

In a precedent paper \cite{Souza-Neto-APL03} we reported on the evidence
of distortion within the coordination octahedra around the Mn in $La_{0.7}Sr_{0.3}MnO_{3}$
thin films from the angular dependence of the XANES (X-ray Absorption
Near Edge Structure) spectra in strained films. In the present paper
this study has been completed by new set of data, a quantitative analysis
of the XANES spectra using the Natoli's rule\cite{Natoli-84} and
of EXAFS (Extended X-ray Absorption Fine Structure) measurements in
the plane of strained and relaxed films. This complete set of experimental
results is combined to \emph{ab initio} calculation of angular resolved
XAS for a local distortion in tensile and compressive substrates,
that account for the angular dependence of the spectra. We found a
tetragonal distortion of the $MnO_{6}$ and no significant modifications
in the $Mn-O-Mn$ angle. These findings stress the importance of the
distortion of the ligand field around the manganese atoms. The strain-induced
modification in the film properties should then be mainly ascribed
to $t_{pd}^{0}$ and the variations of hybridization between the metal
and ligand orbitals, which tends to localize the $e_{g}$ electron
at the $Mn$ site. In the section \ref{sec:X-ray-absorption-spectrocopy}
we present the basis of the angular resolved X-ray Absorption Spectroscopy
and remind the basis of the Natoli's rule and the principles of data
analysis. In section \ref{sec:EXPERIMENTAL} we give the characteristics
of the sample and thoroughly describe the experiments. The results
obtained from XANES and EXAFS analysis are detailed and discussed
in section \ref{sec:Results} and \ref{sec:Discussion}. The prominent
aspects of this study are then summarized in section \ref{sec:Conclusions}.

\section{\label{sec:X-ray-absorption-spectrocopy}X-ray absorption spectroscopy }

The absorption cross section $\sigma(\omega)$, ratio between the
absorbed energy and the incident photon flux, is given by the summation
over all possible final states of the transition probability from
the initial $\mid i>$ to a final state $\mid f>$. In the dipolar
approximation of Fermi's Golden Rule\cite{Brouder-JPCM90}, each probability
can be expressed by $\left|\left\langle f\left|\hat{\epsilon}\cdot\vec{r}\right|i\right\rangle \right|^{2}$.
Due to the dot product $\hat{\epsilon}\cdot\vec{r}$ , the absorption
cross section in anisotropic media has the same structure as the dielectric
constant and can be described by a tensor of rank two, whose expression
depends on the point group of the media\cite{Brouder-JPCM90}. If
the studied system is randomly oriented, the contribution of the dipole
interaction to the absorption coefficient reflects an average of the
contributions in all orientations. In oriented systems, the contribution
of each dipole interaction term can be selected. The simplest expression
of the dichroic effect is obtained for non-cubic samples, with a rotation
axis of order greater than two, where one can find two different cross
sections. For this particular case, and expressing in terms of the
linear absorption coefficient $\mu\propto\sigma(w)$, one defines
two parameters: $\mu_{\Vert}$ stands for the absorption coefficient
when the electric vector lies in a plane orthogonal to the rotation
axis and $\mu_{\bot}$ is the coefficient when the electric vector
is along the rotation axis. For any given orientation of the electric
vector, measured by the angle $\theta$ related to that rotation axis,
$\mu(\theta)$ reads:\begin{equation}
\mu(\theta)=\mu_{\Vert}sin^{2}\theta+\mu_{\bot}cos^{2}\theta\label{eq:mue-dicroism}\end{equation}
This expression is valid over the whole energy XAS range\cite{Brouder-JPCM90}.
Selective pieces of information can be extracted using angle-resolved
XAS, as long as oriented samples are available. Angle resolved XAS
is widely used to improve the results of conventional analysis and
enhances the sensitivity of XAS to probe very tiny difference in anisotropic
systems, like anisotropic single-crystals\cite{Tolentino-PhyC92,Gaudry-PRB03},
surfaces\cite{Magnan-PRL91} and multilayers \cite{Pizzini-PRBRC92}.

The angle resolved EXAFS data have been analyzed following standard
procedures\cite{Koningsberger88}. For the analysis of the XANES data
we combined to \textit{ab initio} simulations, a semi-empirical approach
based on the so called Natoli´s rule\cite{Natoli-84}.

XANES $Mn$ K-edge spectra were calculated by real space full multiple
scattering using the Feff8.2 code \cite{Ankudinov-PRB98}. The potentials
are modeled with the Hedin-Lundqvist exchange correction, taking into
account a single hole in $Mn$ $1s$ orbital, without screening effect.
Atomic positions are given for a set of atoms whose centers are located
at a distance $R_{c}$. The electric field vector polarization is
defined in relation with the atomic structure orientation, and then
all the photoelectron scattering terms, are weighted using the equation
(\ref{eq:mue-dicroism}). This procedure enables to calculate independently
the information from the different angular contributions to the absorption. 

The calculations are conducted self consistently in a sphere of radius
$RSCF$, providing a calculation of the Fermi energy $(k=0)$ accurate
to a couple of eV. Full multiple scattering is taken into account
for $R_{c}<RFMS$, while only single scattering contributions are
contemplated for outer atoms. No specific dynamical contribution to
the disorder (Einstein or Debye model) was introduced in the calculations.
The radial disorder has been modeled by a Gaussian broadening, where
$\sigma$ has been approximated by the Debye-Waller factor obtained
from the EXAFS analysis of the coordination shell $(\sigma=0.06\textrm{Å)}$.
The broadening due of the experimental energy resolution has been
partially taken into account by including a shift of 0.6eV in the
imaginary part of the potential. The calculated spectra were normalized
by the value at 50eV above the absorption edge and a small energy
shift of -1.8 eV has been applied to fit the edge position in the
experimental data. 

\begin{table*}

\caption{\label{tab:samples} Samples nomenclature and characteristics obtained
by X-ray diffraction. Biaxial strain factors can then be defined as
$\varepsilon_{\Vert}=\frac{a_{\Vert}-a_{bulk}}{a_{bulk}}$ and $\varepsilon_{\bot}=\frac{a_{\bot}-a_{bulk}}{a_{bulk}}$,
where $a_{\Vert}$ and $a_{\perp}$ are the in plane and out of plane
lattice parameters of films respectively.}

\begin{tabular}{cc>{\centering}p{90pt}>{\centering}p{90pt}>{\centering}p{90pt}}
\hline 
Composition (nomenclature)&
Strain (relaxation)&
Thickness (relaxation)&
Lattice parameter&
Strain factors ~$\epsilon_{\parallel}$ ($\epsilon_{\perp}$)\tabularnewline
\hline
\hline 
$La_{0.7}Sr_{0.3}MnO_{3}$ (LSMO)&
-&
$\infty$&
$3.87$\AA&
-\tabularnewline
$LaAlO_{3}$ (LAO300)&
Fully relaxed&
300nm ($t_{c}\approx30nm$)&
$3.793$\AA&
$\approx0\%$\tabularnewline
$LaAlO_{3}$ (LAO45)&
Compressive ($\approx10\%$ relaxed)&
45nm ($t_{c}\approx30nm$)&
$3.793$\AA&
-2.0\% (2.3\%)\tabularnewline
$MgO$ (MO60)&
Fully relaxed&
60nm&
$4.21$\AA&
$\approx0\%$\tabularnewline
$SrTiO_{3}$ (STO60)&
Tensile (fully strained)&
60nm ($t_{c}\approx100nm$)&
$3.905$\AA&
0.9\% (-0.8\%)\tabularnewline
\hline
\end{tabular}\end{table*}

The Natoli's rule expresses the linear relationship between the wave
vector of the photoelectron at a multiple scattering resonance, and
$1/R$, where $R$ is the distance to the nearest neighbors. It is
justified theoretically within the framework of the multiple scattering
theory. The absorption cross section is determined by the multiple
scattering matrix $M$ of the photoelectron with kinetic energy $(\hbar k)^{2}/2m=E_{r}-E_{0}+V$,
with $E_{r}$ energy of the resonance, $E_{0}$ absorption threshold
and $V$ average muffin-tin interstitial potential. The maxima of
the absorption correspond to the condition $DetM=0$, giving the relation
$k\cdot R=constant$. The extraction of the variation of the interatomic
distance in unknown systems is complicated by the determination of
$k=\left[\hbar/2m(E_{r}-E_{0}+V)\right]^{1/2}$ because $V$ is unknown
and cannot be determined experimentally. To overcome this problem
the energy separation of a multiple resonance in the continuum from
a bound state at the threshold $\Delta E_{r}=(E_{r}-E_{b})$ can be
used to determine the variation of the interatomic distance\cite{Natoli-84}.
The rule is then expressed under its usual form : \begin{equation}
\left(E_{r}-E_{b}\right)R^{2}=K\label{eq:natolis-rule}\end{equation}
 This rule is valid only for a range of about 20\% variation of the
interatomic distance R, where the energy dependence of the scattering
phase shifts is negligible. It has been successfully applied to the
case of Cu-O bonds in a large series of high temperature superconductors\cite{Tolentino-PhyC92}.
In these systems the constant $K$ was found to be about $46eV.$$\textrm{Å}$$^{2}$.
This corresponds to a $\Delta E_{r}=11.4eV$ for a $Cu-O$ bond of
$1.95\textrm{Å}$\cite{Tolentino-PhyC92}.

\section{\label{sec:EXPERIMENTAL}Experimental}

The $La_{0.7}Sr_{0.3}MnO_{3}$ (LSMO) thin films have been epitaxially
grown in the {[}001{]} direction by pulsed laser deposition under
tensile (SrTiO$_{3}$ {[}001{]}) and compressive (LaAlO$_{3}$ {[}001{]})
substrates with cubic and pseudo-cubic structures. The small lattice
mismatch between LSMO, STO and LAO allows a pseudomorphic growth and
the films are fully constrained for thickness below the critical thickness
($t_{c}$) for each case\cite{Ranno-ASS02,FavreNicolin-PhD03}. MgO
substrate (cubic) with large lattice mismatch (9\%) was used to obtain
an unstrained textured film. A summary of the sample characterization
by X-ray diffraction is given in table \ref{tab:samples}, and more
details can be found elsewhere \cite{Ranno-ASS02,FavreNicolin-PhD03}.

The x-ray absorption experiments were performed at the D04B-XAS beamline
of the Brazilian synchrotron light laboratory (Laboratório Nacional
de Luz Síncrotron, LNLS) in Campinas\cite{Tolentino-JSR01}, Brazil
at the Mn K-edge (6.5keV). The monochromator was a Si (111) channel-cut.
An ion chamber monitored the incident beam. 0.5 mm-slits selected
beam in the orbit plane $(\Psi=0)$ with an acceptance of 0.03mrad.
The light on the sample was predicted to be more than 99\% linearly
polarized\cite{Tolentino-JSR04}. The XANES spectra were collected
in the range 6440 to 6700eV with 0.3eV-energy steps. The instrumental
energy resolution was 1eV, of the same order as the core hole width
(1.16eV), leading to an overall resolution of 1.5eV. During a sequence
of XANES experiments, thermal effects in the optics induce a small
energy shift. This shift was carefully monitored by recurrent collections
of the XANES spectra of a Mn metal foil (edge in 6539.1eV). The energy
scale of the XANES spectra were then corrected using this calibration
curve. Moreover the corrected energy scale was checked by performing
the same operations on second set of XANES data collected in a inversed
time sequence. Energy shifts as small as 0.1eV are certified. The
XANES were normalized at about 150eV above the edge to be compared
in intensity. 

The experimental setup included a goniometer with the rotation axis
perpendicular to the orbit plane. The sample plates with surface of
5$\times$5mm$^{\textrm{2}}$ were placed on the goniometer with the
surface aligned with the goniometer axis. The spectra were collected
in two geometries: with the electric field of the incident linearly
polarized photon beam set approximatively parallel $(\mu_{\Vert})$
and perpendicular $(\mu_{\bot})$ to the film surface. Actually the
working angles were $10^{\circ}$and $75^{\circ}$, that can induce
systematic errors for $\mu_{\parallel}$ and $\mu_{\perp}$. But these
errors are low ( <3\% for $\mu_{\parallel}$ and about 7\% for $\mu_{\perp}$)(eq.
(\ref{eq:mue-dicroism})) \textit{}and have been dismissed\textit{.}
For $\mu_{\parallel}$ measurements the beam size at the sample holder
was $0.1\times4.5mm^{2}$. The lateral size of the beam was reduced
by a factor 4 for $\mu_{\perp}$ measurements.

To check possible errors in the alignment, as well as any kind of
sample anisotropy in the plane, the sample, with a given $\theta$
were also mounted on a goniometer with rotation axis along the propagation
axis of the photons beam. $\mu_{\theta}-$XANES spectra with the same
$\theta$ but rotated around the propagation axis beam were then collected.
The spectra were identical according to the experimental error, confirming
that the samples are isotropic in the plane

Data acquisition were performed in the fluorescence mode collecting
the Mn $K_{\alpha}$ (near 5.90keV) photons using a Ge 15-elements
solid state detector. With this photons energy, about $10\mu m$ of
substrate is probed beyond the whole films (25 to 300nm). This raises
no much trouble for $STO$ substrate, but for $LAO$ substrate, the
$La-L_{\alpha,\beta1,\beta2,\gamma}$ fluorescence lines of the substrate
are more intense, and superimposed to the magnitude of the $Mn-K_{\alpha}$
fluorescence and their contribution cannot be totally dumped by the
selection window. In the $LAO45$ sample, the total acquisition time
reaches a minimum of 4 min per point for the XANES range, and EXAFS
data with satisfactory signal over noise ratio could not be collected.
For the La-free substrates (STO and MO) due to the reduced beam size
and the occurrence of a large number of Bragg peaks in this geometry,
it was not possible to obtain $\mu_{\perp}$ EXAFS data of the required
quality. Only $\mu_{\parallel}$ EXAFS data were collected, over 800eV
above the edge.

\section{\label{sec:Results}Results }

\subsection{$Mn$-K XANES}

\begin{figure}
\begin{center}\includegraphics{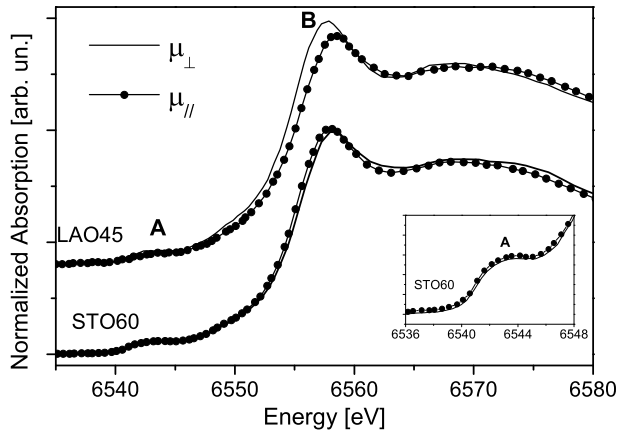}\end{center}

\caption{\label{fig1:ipop_laosto_exp}Experimental XANES spectra for the $\mu_{\perp}$
and $\mu_{\parallel}$ contributions in LAO45 and STO60 films. The
Inset shows the pre-edge structure for STO60 film.}
\end{figure}

The $\mu_{\parallel}$ and $\mu_{\perp}$XANES spectra for tensile
and compressive films (fig. \ref{fig1:ipop_laosto_exp}) show the
classical features observed in manganites compounds\cite{Qian-PRB01_2,Tyson-Mustre-PRB96}:
pre-edge peaks (A) \cite{Croft-PRB97,Qian-PRB01}, arising from p-d
mixing transition when the $O_{h}$ symmetry is broken and relatively
sharp rising edge and broad bond resonances, dominated by the ''white
line'' (labelled resonance B) assigned to transitions to $4p$ empty
states.

In the pre-edge range no difference can be evidenced between the angle
resolved $\mu_{\parallel}$ and $\mu_{\perp}$ (fig. \ref{fig1:ipop_laosto_exp},
insert), or between of the angle resolved XANES of the strained films
and the isotropic spectra of the relaxed ones. 

Above the rising edge two main effects are observed: an energy shift
in the position of the main jump and correlatively in that of the
resonance B, and changes in the shape and amplitude of this resonance.
The shifts for the tensile and compressive films are in opposite directions\textit{.}
For the tensile substrate STO60, the negative energy shift (-0.4eV)
in the in-plane spectrum shows that the average $Mn-O$ bond length
is larger in the film plane than out-of-plane of the film. On the
other hand in the film LAO45 under compressive strain, the positive
energy shift (+0.9eV) indicates that the average bond length is smaller
in plane as out of the plane of the film (Natoli's rule - eq. (\ref{eq:natolis-rule})).
We must emphasize here that as the valence state remain unaltered,
all energy shift effects discussed in this paper are related to bond
length changes.

\begin{figure}
\begin{center}\includegraphics{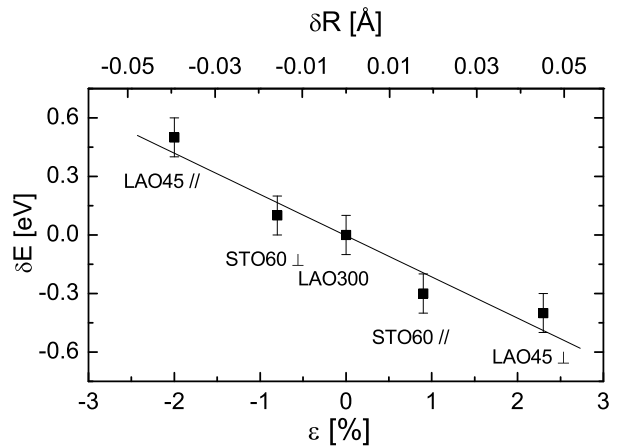}\end{center}

\caption{\label{fig2:dEdR-exp}Energy shift at the Mn K-edge relative to the
relaxed film, plotted against the strain induced by the substrate.
At the top scale, the strain is associated to the local $Mn-O$ bond
lengths. }
\end{figure}

We observe that the amplitude of the shift for the LAO45 film is about
twice that for the STO60 one, and of the same order as the ratio of
the long range strain factor among these films. These results are
summarized in the figure \ref{fig2:dEdR-exp}. The strain on the cell
parameters is directly related to average octahedral modifications
and suggests a model of tetragonal distortion at the atomic scale.
Considering a cluster centered in the manganese, a model structure
is built by transforming the coordinates (x,y,z) of a given atom to
($x+\varepsilon_{\Vert}x,y+\varepsilon_{\Vert}y,z+\varepsilon_{\bot}z)$.
The angle $\varphi$ is not modified. The length of the $Mn-O$ bond
in the plane and out of the plane of the films are proportional to
the corresponding cell parameters and directly derived from the strain
factor : $R_{\Vert}=R_{0}(1+\varepsilon_{\Vert})$ and $R_{\bot}=R_{0}(1+\varepsilon_{\bot})$
.

This model can be confronted with the experimental data using the
Natoli´s rule (see section \ref{sec:X-ray-absorption-spectrocopy})\cite{Natoli-84}:
$\left(E_{r}-E_{b}\right)R^{2}=K$. The resonance considered here
is the B resonance corresponding to $1s\rightarrow4p$ antibonding
$\sigma^{*}$ transition. The energy $E_{r}$ is then the energy of
maximum at the edge. For small changes in the ligand distance $\delta R=R-R_{0}$,
resulting in small energy shifts $\delta E=E_{r}-E_{r0}$, with respect
to the reference energy $E_{r0}$, a linear relationship is obtained
by differentiation: $\delta E=-2\frac{\delta R}{R_{0}}(E_{r0}-E_{b})$.
In the following the relaxed film $LAO300$ is taken as reference
($E_{r0}=6558.0eV$, $R_{0}=1.948\AA$). The experimental energy shift
related to the position of main resonance $\delta E$ can be plotted
as a function the variation of distance with respect to the distance
$R_{0}$, as introduced by the top scale in figure \ref{fig2:dEdR-exp}.
The relation is linear, in accordance with the Natoli's rule, giving
for $\Delta E_{rb}=E_{r0}-E_{b}\approx10.5eV$.

\begin{figure}
\begin{center}\includegraphics[%
  scale=0.7]{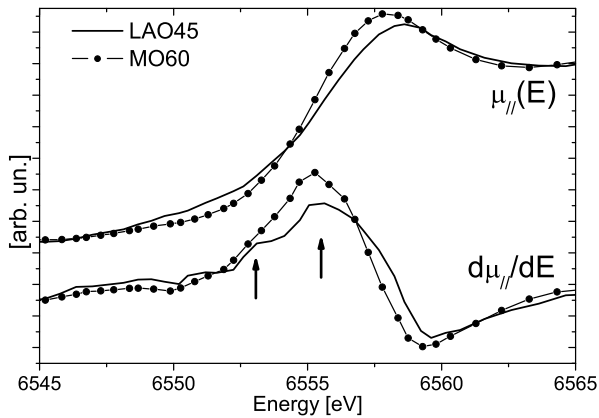}\end{center}

\caption{\label{fig3:inclinacao-da-borda}XANES spectra and its derivative
for in-plane measurements on LAO45 and MO60 films. The vertical arrows
indicate contributions likely coming from two different distances.}
\end{figure}

An additional remark should be made on the comparison of the XANES
spectra in strained and relaxed films: the main slope at the absorption
edge is larger for the strained films than for the relaxed one, and
the derivative (fig. \ref{fig3:inclinacao-da-borda}) reveals the
presence of two contributions (peaks marked by vertical arrows). This
result will be discussed in the next section, along with the EXAFS
results. 

The modifications of the main line are not straightforward interpreted
by qualitative considerations and we resort to \textit{ab initio}
simulations.  The clusters used in calculations are build from the
structure and cell parameters of $La_{0.7}Sr_{0.3}MnO_{3}$ regular
octahedron $MnO_{6}$ in the symmetric $D_{3d}^{6}$ group, and considering
only $La$ in the $A$ site. Identical calculations were performed
with several random combinations of $Sr$ and $La$ on this site.
These different calculations lead to similar spectra. 21 atoms clusters
calculations (central $Mn$, 6 $O$ in the coordination shell, 8 $(La,Sr)$
atoms and the third shell of 6 $Mn$ atoms) give a good agreement
with main features of the experimental spectra, relevant for the present
study. As a matter of fact the main features discussed here are determined
by the coordination polyhedron. Thence for the simulations of the
strained structures only the coordination polyhedra were distorted,
and the $A$ atoms and the $Mn$ shell are kept in the positions of
the relaxed structure, using local order parameters scaling with the
crystallographic cell parameters of the films. The 21-atom spectra
correspond to a radius $R_{c}=4\AA$ (all calculations use $RFMS=4\AA$
and $RSCF=4\AA$). 

\begin{figure}
\begin{center}\includegraphics{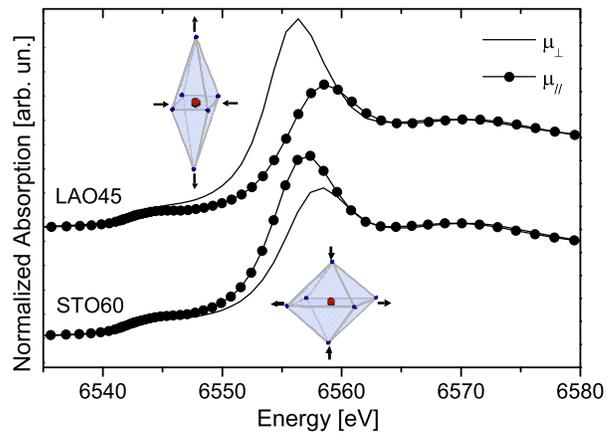}\end{center}

\caption{\label{fig4:ipop_laosto_sim}Simulated XANES spectra for the $\mu_{\perp}$
and $\mu_{\parallel}$ contribution of the supposed LAO45 and STO60
films structures. }
\end{figure}

The simulations were done using two models for $MnO_{6}$: keeping
the $O_{h}$ symmetry for $MnO_{6}$ or considering a tetragonal distortion.
As expected for this isotropic case, $\mu_{\parallel}$ and $\mu_{\perp}$
XANES spectra are identical. We also observe that in simulated spectra
keeping the $O_{h}$ symmetry for $MnO_{6}$ the shape of the of the
B feature is almost unaltered. Thence a remotion of the octahedra
symmetry is necessary to account for the alterations in the XANES
spectra of the strained films. The calculations shown in figure \ref{fig4:ipop_laosto_sim}
were performed for clusters with tetragonal distortion of the $MnO_{6}$
{}``octahedra''. The calculated structures reproduce well the main
features of the experimental results. They account as well for the
energy shift, in amplitude and direction, as for the relative reduction
of the amplitude of the main feature close to the edge, among the
two orientations for each film (fig. \ref{fig1:ipop_laosto_exp}).

\subsection{$Mn$-K EXAFS}

\begin{figure}
\begin{center}\includegraphics{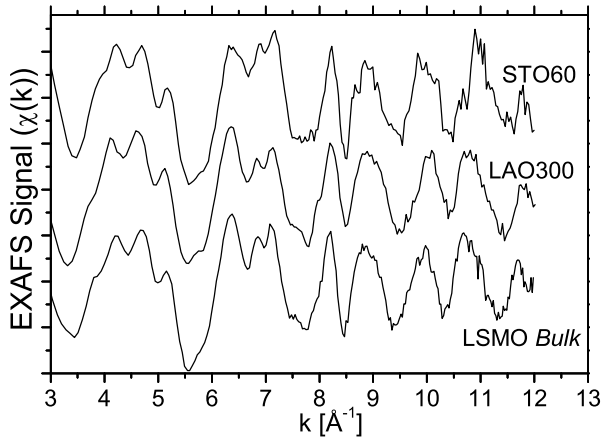}\end{center}

\caption{\label{fig5:chi}Mn K-edge EXAFS measurements of the films and bulk
$La_{0.7}Sr_{0.3}MnO_{3}$.}
\end{figure}

As EXAFS signal cannot be collected for LAO45, our study is limited
to the STO60 sample. A complete EXAFS characterization of the octahedral
deformation would require a measure of the $Mn-O$ out of plane distance
from the $\mu_{\bot}$- EXAFS in this sample. As we mentioned in the
section \ref{sec:X-ray-absorption-spectrocopy} for $\mu_{\bot}$measurements
the presence of a large number of Bragg peaks associated to a lower
flux prevents from the collection of data with proper quality for
this analysis. EXAFS signal for $La_{0.7}Sr_{0.3}MnO_{3}$ bulk and
$\mu_{\parallel}$ in strained and relaxed films, are shown in figure
\ref{fig5:chi}. The extraction of this signal is a critical point
when one intends to compare data collected with non-equivalent and
relatively low signal over noise ratio. To certify the results we
resort to several codes \cite{Klementev-JPD01,Michalowics-PhD90,Ravel-PRB99,Ressler-JP97}
making use of different extraction procedures based on spline or polynomial
fits of the atomic background or/and Fourier transform filtering.
The tail of the $La$ $L_{1}$ XAS spectra $(E_{0}=6267eV)$ is largely
removed by the signal extraction. It contributes weakly to low frequency
component in the EXAFS spectra of the bulk $La_{0.7}Sr_{0.3}MnO_{3}$.
Small departure from the bulk-like correction in the films have been
neglected. The extension of the EXAFS spectra is limited in the films
limiting the $r$-space resolution of the study through the relation
$\delta r.\delta k\approx\pi/2$. However, a discussion of the local
distortion in $La_{0.7}Sr_{0.3}MnO_{3}$ bulk is out of the scope
of this paper. The information we seek here is related to small shifts
with respect to a relaxed bulk-like situation, whose spectra are then
taken as reference. 

\begin{figure}
\begin{center}\includegraphics{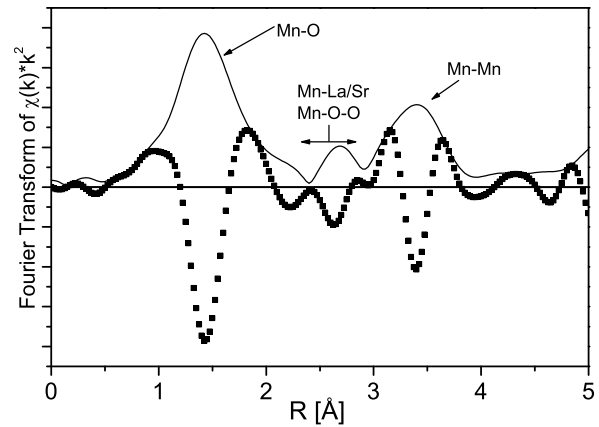}\end{center}

\caption{\label{fig6:FT-STO}Fourier transform of the EXAFS signal for the
tensile film (STO60). The FT modulus (solid line) and the FT imaginary
part (points) are shown.}
\end{figure}

The Fourier transform of the EXAFS signal of $La_{0.7}Sr_{0.3}MnO_{3}$
bulk is given in the figure \ref{fig6:FT-STO}. The first main peak
corresponds to the oxygen coordination shell. In our low resolution
data this shell can be properly fitted using calculated amplitude
and phases relative of the Mn-O pair on the basis of the crystallographic
structure, by assuming one Mn-O distance. According to our comparative
approach, we extracted from the signal of this coordination shell,
amplitude and phases relative to the Mn-O pairs, that will be later
used to refine the data of the films. At higher distance in the $FT$
are shown the peaks corresponding to larger effective distances, including
the A site neighbours and nearest-Mn single scattering (SS) contributions,
as well as the multiple scattering (MS) contributions. In manganite
compounds all these contributions must be taken into account together.
In the present study we resort to \emph{ab initio} simulations on
the basis of the crystallographic structure to evaluate how far the
analysis of the Mn next nearest neighbours can be simplified. Due
to the high symmetry of atomic distribution around $Mn$ atoms, the
contribution of this shell is included in the relatively sharp FT
peak at around 3.5 $\textrm{Å}$. The simulations reveal that so far
the low k part $(k\leq4.5\textrm{Å}^{-1})$ of the signal and the
peak sides and are not included (back Fourier Transform reduced to
3.3-4 $\textrm{Å}$), the MS is dominated by the contribution the
almost collinear of 4-fold paths involving Mn nearest neighbours and
their common oxygen. We observe that this path is closely related
to the Mn-Mn SS path, because Mn-Mn is one of its legs. The two other
legs are Mn-O legs with a focusing angle close to 180, resulting to
an effective length of the path very close to Mn-Mn distance. The
length variation of the SS and MS dominant paths are then closely
related. A variation in the Mn-Mn bond lengths will result, in a good
approximation, to an identical variation in the effective length of
the MS path. A new MS path can be parameterized constraining its parameters
to be the same as those of the SS path. It was verified that a fitting
procedure taking into account the two contributions with the same
fitting parameters gives a satisfactory sensitivity to a variation
of the Mn-Mn distance. Our fits have then been performed in this simple
way, using both SS and MS paths, but without adding new variables
to the fit, in compatibility with the limited number of independent
points.

The signal of first peak of the Fourier transform in relaxed and strain
films was back-transformed over the $R$-range 0.9 to 2.04$\textrm{Å}$
for the fitting procedure. We used relative phase and amplitude data
obtained by two ways: extracting from experimental data for bulk material
and from \emph{ab initio} simulations. Identical results are obtained
in the two cases. In the relaxed film MO60 (or LA300) a single shell
fit with $Mn-O$ bond of $(1.95\pm0.02)\AA$~accounts for the experimental
data. In the strained film STO60, we obtained for the bond length
a larger average value $\left\langle Mn-O\right\rangle $ of $(2.01\pm0.02)\AA$.
In this case, we should point out that the achievement of repetitive
fits of equivalent quality, needed the introduction of two different
$Mn-O$ bonds at $(1.94\pm0.02)\AA$ and $(2.05\pm0.02)\AA$ respectively
(table \ref{tab:resultados-exafs}). One-shell fits lead to non-repetitive
or meaningless results and quality factors worse by at least a factor
4. 

\begin{table}

\caption{\label{tab:resultados-exafs}EXAFS results, for relaxed film (MgO)
and under in-plane expansion (STO). Considering only the in-plane
contribution.}

\begin{center}\begin{tabular}{m{30pt}>{\centering}m{40pt}>{\centering}m{50pt}>{\centering}m{48pt}>{\centering}m{48pt}}
\hline 
\vspace*{3pt}
~\vspace*{3pt}
&
\multicolumn{4}{c}{\textbf{Coordination shell ($Mn-O$)}}\tabularnewline
\cline{2-5} 
\hline 
~&
N&
R (\AA) &
$\sigma^{2}\cdot10^{4}$(\AA$^{2}$)&
$\Delta E_{0}$(eV) \tabularnewline
\vspace*{3pt}
STO60\vspace*{3pt}
&
$1.3\pm0.5$

$2.7\pm0.5$&
$1.94\pm0.02$

$2.05\pm0.02$&
$17\pm5$

$17\pm5$&
$-1.4\pm0.2$

$-1.4\pm0.2$\tabularnewline
\vspace*{3pt}
MO60&
$4$(fixed)&
$1.95\pm0.01$&
$25\pm5$&
$-1.4\pm0.2$\tabularnewline
\hline
\hline 
\vspace*{6pt}
\vspace*{6pt}
&
\multicolumn{4}{c}{\textbf{$Mn$ next neighboring shell ($Mn-Mn$)}}\tabularnewline
\hline
~&
N&
R(\AA) &
$\sigma^{2}\cdot10^{4}$(\AA$^{2}$)&
$\Delta E_{0}$(eV)\tabularnewline
\vspace*{3pt}
STO60\vspace*{3pt}
&
$7\pm1$&
$3.890\pm0.005$&
$25\pm5$&
$0.4\pm0.2$\tabularnewline
\vspace*{3pt}
MO60&
$6\pm1$&
$3.875\pm0.005$&
$13\pm5$&
$0.4\pm0.2$\tabularnewline
\hline
\end{tabular}\end{center}
\end{table}

The signal of Mn nearest neighbors has been selected by back-Fourier
transform in the R-range $3.3\AA$ to $3.95\AA$. The low range part
(< 4.5\AA$^{-1}$), was dismissed. Procedures using backscattering
amplitude and phases obtained from a bulk sample give good fits and
the $Mn-Mn$ distances are determined with small relative uncertainties
($\sim$0.005\AA$^{-1}$). The relaxed films the $Mn-Mn$ distance
is found to be $3.875\AA$, as it is in the bulk compound. In the
tensile STO60 film this distance is $3.89\AA$, i.e. increased by
about $0.02\AA$ in comparison to bulk compound and the relaxed films
(table \ref{tab:resultados-exafs}).

\section{\label{sec:Discussion}Discussion}

From $\mu_{\Vert}$- EXAFS we found an average $Mn-O$ in-plane distance
larger in the strained STO60 film as compared to the relaxed samples.
This states for an enlargement of the basal $MnO_{4}$ square. Much
care should be taken, however, on the interpretation of the numerical
values associated to these fits. In $La_{0.7}Sr_{0.3}MnO_{3}$ bulk
compounds it has been observed that the local structure around Mn
atoms can be described as well by a model of regular octahedron or
a distorted octahedron with different Mn-O distances, these models
cannot be unambiguously distinguished even in a high resolution study\cite{Shibata-PRB03}.
In the present low resolution study the absolute value of the variation
in $Mn-O$ distance between strained and relaxed film ($0.06\pm0.04Å$)
has a limited significance, and only express the expansion of this
distance, in agreement with the XANES analysis. 

The $\mu_{\Vert}$- EXAFS analysis of the coordination shell, however,
rises another comment. It is well settled that in an EXAFS experiment
the k-range available limits the minimum distance separation $\Delta R$
that can be resolved in the analysis. It is usually considered that
$2\Delta k\Delta R$ should be larger that $\pi$. From this criteria
distance separation lower than 0.12$\textrm{Å}$ cannot be unambiguously
evidenced in our study. However we consider that the tendency to obtain
two separate sub-shells for the fit of the coordination shell in the
constrained film STO60 should be noted. We should stress that our
approach is comparative and that, when a two-shell fit is performed
with the same conditions in the relaxed samples, the two distances
invariably collapse to give one unique value. This indication of the
possible coexistence of two distances in the plane has to be related
to another experimental evidence obtained from the study of the rising
edge in constrained films, where two contributions were observed (fig.\ref{fig3:inclinacao-da-borda}).
These observations can lead to two different conclusions. In a first
hand the biaxial strain may induce an additional unexpected small
additional distortion of the basal plane of the $MnO_{6}$ octahedra.
This distortion would express a tendency to recover a Jahn-Teller
distortion in this plane. On the other hand the observation of two
distances can be related to the existence of two domains with different
relaxation states, as observed by Qian et al.\cite{Qian-PRB01_2}
in $Nd_{0.5}Sr_{0.5}MnO_{3}$. \textcolor{green}{}\textcolor{black}{Depth
sensitive local probe as X-ray absorption measurements at grazing
angle would be necessary to settle this point}\textcolor{green}{.}
No quantitative evaluation of this -clearly very small- splitting
can be given and it will be dismissed in the following. 

Based on the differential Natoli's rule, $\delta E=-2\frac{\delta R}{R_{0}}(E_{r}-E_{b})$,
and on the model of tetragonal distortion for the octahedra reflecting
the film strain, we deduce from the slope of the curve in figure \ref{fig2:dEdR-exp}
$\Delta E_{rb}=(E_{r}-E_{b})=10.5\pm0.1eV$ . Using the $Mn-O$ distance
of $1.95\textrm{Å}$ of the relaxed films, a constant $K$ of about
$41eV.$$\textrm{Å}$$^{2}$ is found. This value is close to that
obtained for Cu-O bonds in high temperature superconductors\cite{Tolentino-PhyC92}.
Due to the electronic similarities between these two metallic perovskite
systems, the interstitial potentials are expected to be similar. The
value of $\Delta E_{rb}$ fix the position of the bond level $E_{b}$
in the XANES spectra. The energy of the resonance B is $6558eV$ in
the relaxed spectra, so that $E_{b}=6547eV,$ about $4eV$ above the
pre-edge structure (A) and close to the onset of the main edge (fig.\ref{fig1:ipop_laosto_exp}).
It is worth noting that the actual position of the bond level for
Cu-O in high temperature cuprates is also few eV apart from the pre-edge
features \cite{Tolentino-PhyC92}. The agreement between constant
K and the reasonable value for the bonding energy express the validity
of the model. We consider that for the coordination shell the values
determined by XANES analysis far more precise than those obtained
from EXAFS and further discussion is based on these values.

The first important statement deduced from our XANES results is that
$MnO_{6}$ geometry is directly affected by the strain and cannot
then be considered as fully determined by the stoichiometry of the
compound. This statement is in agreement with the findings of Qian
and coworkers\cite{Qian-PRB01_2} in a quite different system and
could then be generalized in manganite films. These findings stress
the importance of the distortion of the ligand field around the manganese
atoms. 

In the bulk LSMO system, the $MnO_{6}$ octahedra can be consider
as regular. In the thin films the distortion of the $MnO_{6}$ octahedra
induced by the strain can be expressed by $\sigma=\sqrt{\frac{1}{3}{\displaystyle {\textstyle {\scriptstyle {\textstyle \sum_{i}\left[(Mn-O)_{i}-\left\langle Mn-O\right\rangle \right]^{2}}}}}}$,
in a similar way that for Jahn-Teller distortion. Using the values
deduced from XANES analysis we have $\sigma=0.041\AA$ for the LAO45
film, about one third of the Jahn-Teller distortion in $LaMnO_{3}$.
The trapping of the $e_{g}$ electrons at the $Mn$ site by such tetragonal
{}``Jahn-Teller like'' distortion may largely be responsible for
the lowering of the Curie temperature observed in these films. The
strain-induced modification in the film magnetic and transport properties
should be ascribed to the variations of hybridization between the
metal and ligand orbitals and a reduced transfer interaction $t_{pd}^{0}$,
reducing the double exchange by localizing the $e_{g}$ electron.

The increase of the $Mn-Mn$ distance is determined by EXAFS analysis
with a good accuracy. In the tensile STO60 film this distance is increased
by $0.02\AA$ in comparison to bulk compound and the relaxed films.
This corresponds to a relative variation of 1\%, matching the cell
parameters and of the order of the increase of the $Mn-O$ distance,
as found by XANES analysis. Consequently the $\varphi$ angle is not
-or very weakly- modified. This results are conflicting with those
obtained by Miniotas et al.\cite{Miniotas-JAP01} in $La_{x}Ca_{1-x}MnO_{3}$
films. In almost equivalent strain conditions, they reported variations
of about $8^{\circ}$ in the $\varphi$ angle, and no variation in
the $Mn-O$ distance. They consequently conclude that the biaxial
strain is accommodated by appropriated modification of the tilt angle.
The modification of the $Mn-O$ distance is unambiguously shown by
our XANES data. \textcolor{green}{}Due to the accuracy on the distance
determination and the large angles involved (around $165^{\circ}$)
we cannot conclude about possible small variations (<$2-3^{\circ}$)
of the tilt angle. A limited variation in the tilt angle $\varphi$
may indeed exist. \textcolor{black}{However we have shown that a model
of accommodation at the atomic scale of the strain experienced by
the cell, accounts perfectly for the XANES data, so that tilt angle
variation is not necessary to this accommodation.} Hence we believe
that, in spite of the well established theoretical background and
experimental evidence of the influence of the tilt angle on the magnetic
properties in bulk manganites, the tilt angle is neither the only
nor the main driving factor of the modification of these properties
in strained manganite films.

\section{\label{sec:Conclusions}Conclusions }

We presented here a characterization by an angle-resolved X-ray absorption
spectroscopy, of the local scale structural distortion induced by
substrate epitaxial strain around manganese atoms in $La_{0.7}Sr_{0.3}MnO_{3}$
films. We show that biaxial strain is locally accommodated in the
coordination shell, by distortion of the $MnO_{6}$ octahedron, without
change in the tilt angle. These findings refuse the scenarios where
$MnO_{6}$ octahedron would be entirely determined by the stoichiometry
of the compound and a tilt in the octahedral linkage would be the
only driving parameter in the strain dependence of the transport and
magnetic properties of manganites thin films. We believe that changes
within the octahedra, with an induced tetragonal {}``Jahn-Teller
like'' distortion, should have also a predominant role in the modification
of these properties. 

\begin{acknowledgments}
The authors would like to thank Dr. J. Mustre de Leon for fruitful
discussions. This work is partially supported by LNLS/ABTLuS/MCT and
FAPESP (1999/12330-6). NMSN and AYR acknowledges the grants from CAPES
and CNPq respectively.
\end{acknowledgments}
\bibliographystyle{apsrev}

\end{document}